\documentclass[onecolumn,showpacs,superscriptaddress,prb]{revtex4-1}
\usepackage{graphicx}
\usepackage{textcomp}
\usepackage{amsmath,amssymb,amsfonts,amsthm,latexsym} 
\usepackage[colorlinks=true,urlcolor=blue,linkcolor=blue,citecolor=blue,a4paper]{hyperref}
\usepackage{subfig}
\usepackage{gensymb}
\usepackage{setspace}
 \makeatletter
    \renewcommand\@make@capt@title[2]{%
     \@ifx@empty\float@link{\@firstofone}{\expandafter\href\expandafter{\float@link}}%
      {\textbf{#1}}\@caption@fignum@sep#2\quad}%
    \makeatother
\usepackage[labelsep=quad,justification=raggedright]{caption}

\usepackage{color} 
\usepackage[tableposition=top]{caption}
\begin{document}
\title{Bulk contribution to magnetotransport properties of low defect-density Bi$_2$Te$_3$ topological insulator thin films}
\author{P. Ngabonziza}
\affiliation{Faculty of Science and Technology and MESA+ Institute for Nanotechnology, University of Twente, 7500 AE Enschede, The Netherlands}
\affiliation{Max-Planck-Institute for Solid State Research, 70569 Stuttgart, Germany}
\affiliation{Department of Physics, University of Johannesburg,\\ P.O. Box 524 Auckland Park 2006, Johannesburg, South Africa}
 \author{Y. Wang}
 \affiliation{Max-Planck-Institute for Solid State Research, 70569 Stuttgart, Germany}
\author{A. Brinkman}
\affiliation{Faculty of Science and Technology and MESA+ Institute for Nanotechnology, University of Twente, 7500 AE Enschede, The Netherlands}
\date{\today}
\begin{abstract}
An important challenge in the field of topological materials is to carefully disentangle the electronic transport contribution of the topological surface states from that of the bulk. For Bi$_2$Te$_3$ topological insulator samples, bulk single crystals and thin films exposed to air during fabrication processes are known to be bulk conducting, with the chemical potential in the bulk conduction band. For Bi$_2$Te$_3$ thin films grown by molecular beam epitaxy, we combine structural characterization (transmission electron microscopy), chemical surface analysis as function of time (x-ray photoelectron spectroscopy) and magnetotransport analysis to understand the low defect density and record high bulk electron mobility once charge is doped into the bulk by surface degradation. Carrier densities and electronic mobilities extracted from the Hall effect and the quantum oscillations are consistent and reveal a large bulk carrier mobility. Because of the cylindrical shape of the bulk Fermi surface, the angle dependence of the bulk magnetoresistance oscillations is two-dimensional in nature. 
\end{abstract}
\maketitle
%-------------------------------------------------------------------------------------

%----------------------------------------------------------------------------

\section{Introduction}\label{section_3_1}
Topological insulators (TIs) form a class of semiconductors with an inverted band order due to strong spin-orbit coupling. When the Fermi energy lies in the bulk semiconductor band gap, the bulk of a TI is insulating. But the edges or surfaces host non-gapped states which cross the bulk band gap, providing conduction on the boundaries of the material~\cite{CLKane_2005}. The surface states of a TI mimic relativistic Dirac electrons because of their linear energy-momentum $E(k)$ relation~\cite{LFu_2007}. In the past decade, TIs have received considerable attention since they open opportinuties for exploring a variety of new phenomena in physics~\cite{CLKane_2005,LFu_2007,CWJBeenakker_2013}. Possible fields of application are spintronics and topological quantum computing.
\newline 
\newline
The material Bi$_2$Te$_3$ is one of the second generation 3D TI materials (Bi$_2$Te$_3$, Bi$_2$Se$_3$, and Sb$_2$Te$_3$) investigated extensively~\cite{YYLi_2010}. This material is known to be less prone to forming vacancies compared, for example, to the prototypical Bi$_2$Se$_3$~\cite{MBrahlek_2015,NBansal_2012,PDCking_2014,MBianchi_2010}. In single crystals, a high bulk carrier density is the result of a large defect density. The defects and vacancies act as scattering centers too. An increased carrier density thereby results in a lower bulk mobility~\cite{Ren2010,DXQu2010_Science,KShrestha_2014PRB}.
Recently, using the molecular beam epitaxy technique, progress was made in synthesizing high-quality bulk-insulating Bi$_2$Te$_3$ thin films with very low intrinsic doping~\cite{PNgabonziza_2015,PNgabonziza_2016}. The as-grown material features a Fermi level in the bulk band gap, which is achieved without resorting to techniques such as counterdoping~\cite{TChen_2014,MBrahlek_2014}, p-n layer growth \cite{MLanius_2016,MEschbach_2016,JWang_2012}, or using (off-)stoichiometric ternary~\cite{JZhang_2011} and quarternary~\cite{YXu_2014,TArakane_2012} compounds. 
\newline 
\newline
Besides the successful use of surface-sensitive techniques, such as angle-resolved photoemission spectroscopy (ARPES)~\cite{PNgabonziza_2015,KHoefer_2015} and scanning tunneling microscopy (STM)~\cite{PNgabonziza_2015,ZAlpichshev_2010} to probe the conducting surface states of TIs, studying transport properties in an externally applied magnetic field is also a powerful probe of the nature of the metallic state in TIs~\cite{MVeldhorst_2013}. For example, the characteristics quantum Shubnikov-de Haas (SdH) oscillations can be used to probe the existence of topologically protected surface states; and their analysis reveals as well the existence of additional topologically trivial states in transport characteristics. The SdH effect probes extrema in the cross section of the Fermi surface (FS), and their angular dependence in tilted magnetic fields provides information about the size and the shape of the FS~\cite{ELahoud_2013} and, more importantly, about the dimensionality of the FS~ \cite{AATaskin_2012,LHBao2012}. Early studies on various 3D TI materials have measured SdH oscillations and they were often attributed to originate from the top and bottom topological surface state with the expected Berry phase and angular dependence~\cite{MVeldhorst2012,KWang_2013,Petrushevsky_2012, Devidas_2014, CZhang_2014}. 
\newline 
\newline
Here, we combine structural characterization (transmission electron microscopy), chemical surface analysis as function of time (x-ray photoelectron spectroscopy) and magnetotransport analysis to understand the low defect density and record high bulk electron mobility of Bi$_2$Te$_3$ thin films once charge is doped into the bulk by surface degradation. We caution that two-dimensional Shubnikov de Haas quantum oscillations are often used as evidence for surface state transport, but a careful analysis of low-defect density films reveals that the oscillations might also arise from the high mobility bulk electrons. Using the dimensionality of the bulk Fermi surface, a consistent picture is obtained for the bulk carrier density and mobility as deduced from the Hall effect and magnetoresistance oscillations.

\section{Sample Preparation and Characterization}
\begin{figure*}[!t]
\centering
\includegraphics[width=1.\textwidth]{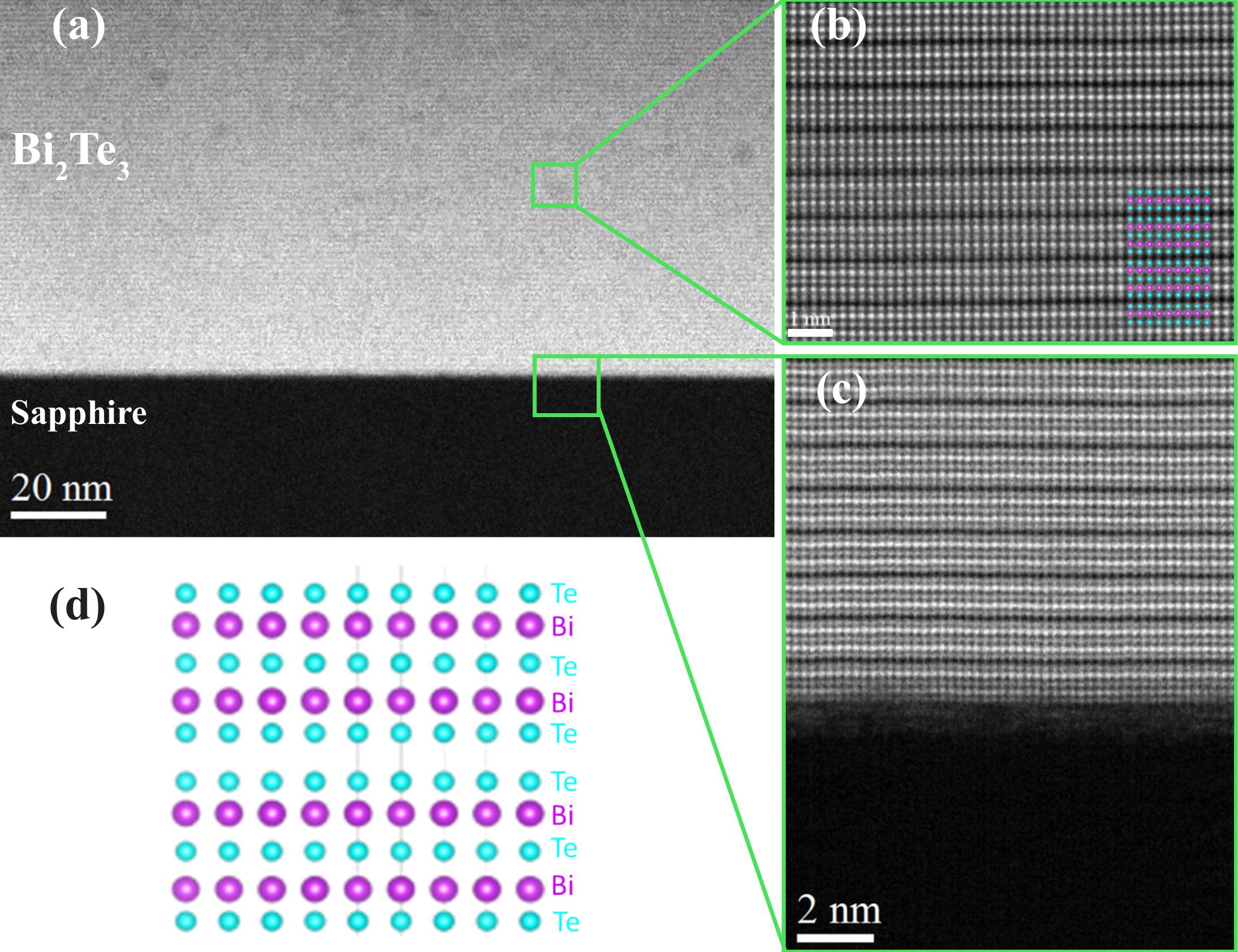}
\caption{\linespread{1} (a) An overview STEM image of a Bi$_2$Te$_3$ film grown on Al$_2$O$_3 (0001)$ substrate. Magnified STEM images of the sample showing: (b) the quintuple layers of the film and (c) a smooth interface between the film and the substrate. (d) Schematic atomic structural model, displaying the sequence of Te and Bi atoms in a quintuple layer, as illustrated in inset of (b).}
 \label{fig:TEM}
\end{figure*}

High quality topological insulator thin films of Bi$_2$Te$_3$ were grown by molecular beam epitaxy (MBE) on Al$_2$O$_3$ [0001] substrates. The base pressure of the MBE chamber was lower than $5\times 10^{-10}$ mbar and the highest pressure recorded during growth was less than $3\times 10^{-8}$ mbar. We employed a two step temperature growth procedure, which results in atomically sharp interfaces between the TI film and the substrate. Details on the growth procedure of our thin films are reported in Refs. \cite{PNgabonziza_2015,PNgabonziza_2016}.
\newline 
\newline
In addition to employing in-situ characterization techniques as described in Refs. \cite{PNgabonziza_2015,PNgabonziza_2016,PNgabonziza_2018}, we used scanning transmission electron microscope (STEM) for a microstructural analysis of the samples presented in this study. STEM investigations were performed on structured Hall bar devices, several months after magnetotransport experiments. Figure~\ref{fig:TEM}\textcolor{blue}{(a)} exhibits a cross-sectional STEM image of a 70 nm Bi$_2$Te$_3$ film grown on sapphire. The image distinctly outlines the film (bright contrast) and the substrate (darker region).
Figure~\ref{fig:TEM}\textcolor{blue}{(b)} shows  a magnified STEM image of the atomic ordering of Bi$_2$Te$_3$ thin film.
Highly parallel quintuple layers are clearly visible in the Bi$_2$Te$_3$ film, which are separated by van der Waals gaps.  The Bi atomic columns appear brightest due to the much higher atomic number of Bi compared to Te, as also indicated in inset with a superimposed structural model (see illustration in Fig.~\ref{fig:TEM}\textcolor{blue}{(d)}). A closer view of the interfacial region between the film and the substrate is shown in Fig.~\ref{fig:TEM}\textcolor{blue}{(c)}. 
Despite large lattice mismatch between Bi$_2$Te$_3  [001]$  and the substrate ($\sim 8.7\%$)~\cite{LHe_2013}, highly parallel layers are clearly visible in the film, thus suggesting a high crystal quality close to TI-substrate interface. Furthermore, no dislocations and distortions of atomic column were observed immediately above the substrate, which confirms rapid relaxation of strain at the interface as previously reported~\cite{XLiu_2011}. This is due to the van der Waals epitaxy, which is known to relax  the lattice-matching condition necessary for most common epitaxial deposition of covalent semiconductors and their heterostructures~\cite{LHe_2013,AKoma_1992}. Although our TEM analysis represents an average of atomic site occupancies over multiple columns, defects are expected to show up. Within our resolution, no Te vacancies could be observed, indicating the high structural film quality that can be attained during film growth under sufficient Te surplus conditions.
\begin{figure*}[!t]
\centering
\includegraphics[width=1.\textwidth]{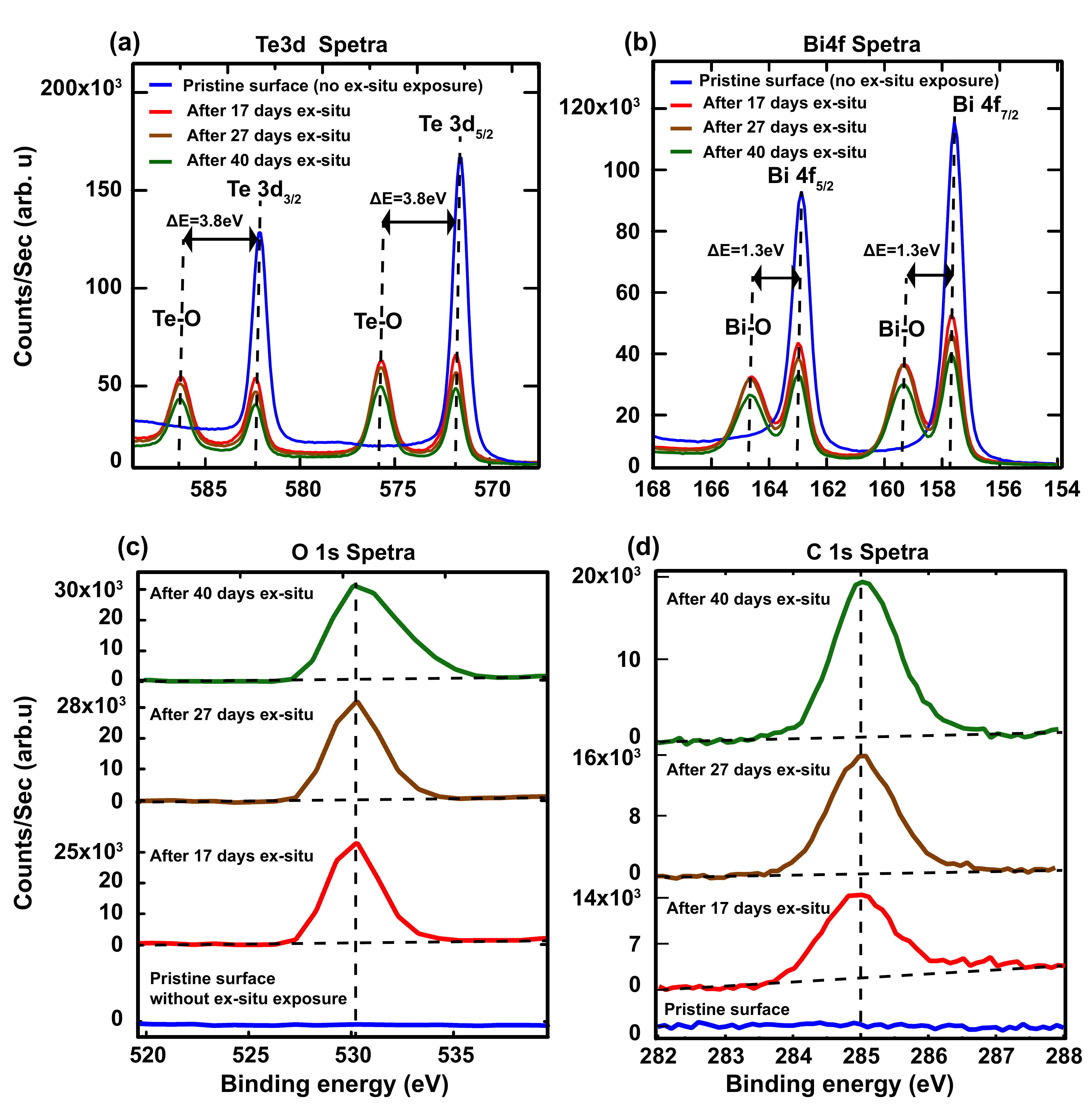}
\caption{\linespread{1.5}Time dependent XPS spectra of a 70 nm Bi$_2$Te$_3$ film grown on sapphire. The spectra were acquired at different time-intervals, before and after exposing the sample surface to air. High-resolution scans around: (a) the Te 3d and (b) Bi 4f main peaks. (c) The O 1s  and (d) C 1s  XPS spectra for the Bi$_2$Te$_3$ surface measured at different exposure times.}
 \label{fig:XPS}
\end{figure*}
\newline 
\newline
The surface elemental characterization and chemical stoichiometry were investigated using X-ray photoemission spectroscopy (XPS), following the procedure presented in Ref.~\cite{PNgabonziza_2015,PNgabonziza_2018}. Through the analysis of the high resolution scans around the Te 3d and Bi 4f peaks for the pristine film, the surface chemical stoichiometry (Te:Bi ratio) was determined to be $1.498\pm0.05$. Since magnetotransport studies and fabrication processes were carried out in ambient conditions, we also examine the surface changes once samples have been exposed to air. For the purpose of keeping track of the contamination process, we performed several XPS measurements, first after sample growth without breaking ultra high vacuum conditions (the XPS chamber is connected to the deposition system via a high vacuum distribution chamber), and later at different time intervals after exposing the sample to ambient conditions. The XPS surface characterizations were performed before structuring samples into Hall bar devices and magnetotransport experiments. 

Figure~\ref{fig:XPS} depicts series of XPS spectra of a 70 nm film acquired at different time-intervals, before and after exposing the sample surface to ambient conditions. These spectra show the aging of the Bi$_2$Te$_3$ surface and provide information about surface elemental composition through the analysis Te 3d, Bi 4f, O 1s and C 1s spectra. The XPS spectra for other films show similar time-dependent contamination behaviour~\cite{PNgabonziza_2015}.
Table~\ref{table_1} gives the binding energies of various peaks measured in XPS before and after exposing the sample surface to air. For the pristine surface, due to spin orbital splitting,  we only observe a pair of peaks for the Te 3d spectra: Te 3d$_{3/2}$ and Te 3d$_{5/2}$ separated by  $\sim 10.3 $ eV; and Bi 4f spectra: Bi 4f$_{5/2}$ and Bi 4f$_{7/2}$ separated by $\sim 5.3$ eV (see Fig.~\ref{fig:XPS}\textcolor{blue}{(a)}-\textcolor{blue}{(b)}). 
This pristine surface shows no signs of oxidation or contaminations, which is confirmed by a flat O 1s and C 1s regions (see Fig.~\ref{fig:XPS}\textcolor{blue}{(c)}-\textcolor{blue}{(d)}); and the absence of higher binding energy peaks in the Te 3d and Bi 4f spectra. After the sample is exposed to air and exposure time increased, there appear new peaks with one type of chemical shift both in the Te 3d ($\Delta E=3.8$ eV) and Bi 4f ($\Delta E=1.3$ eV) spectra. 
This shows that there is formation of an oxidized layer at the film surface as evidenced by the sharp O 1s peak, which shows up at a binding energies of 530.2 eV. The C 1s peak is usually observed in XPS measurements when the sample surface have been in contact with ambient conditions, and they are often attributed to surface contaminations or adsorption of carbonoxides on the surface of the sample~\cite{HBando_2000,AJGreen_2016}. These XPS observations are consistent with previously published data on the exposed surface of Bi$_2$Te$_3$ samples~\cite{PNgabonziza_2015,HBando_2000,BVRChowdari_1996,TPDebies_1977}.
\newline 
\newline
We know that the exposure of the surface of Bi$_2$Te$_3$ thin films to ambient conditions effectively dopes the material with electrons, by which the Fermi energy shifts from the bulk band gap into the bulk conduction band \cite{PNgabonziza_2015}. Below, we will investigate the influence of these bulk carriers on magnetotransport.
\begin{table}[!t]
\centering
\begin{tabular}{|p{2.5cm}|p{2.2 cm}|p{3 cm}|p{ 7cm}|}
\hline
 \textbf{Spectra } & \textbf{Pristine Film [eV]}&\textbf{Oxide \newline Over-layer [eV]  }& \textbf{Literature values [eV]} \\ \hline
 &&&\underline{TeO$_2$} \\ Te 3d$_{3/2}$ &  582.4 & 586.2& 586.2~\cite{HBando_2000}   \\ 
 Te 3d$_{5/2}$ & 572.1&575.9 & 575.8~\cite{HBando_2000} and 575.8~\cite{BVRChowdari_1996}  \\ \hline
 &&&\underline{Bi$_2$O$_3$} \\
 Bi 4f$_{5/2}$&  162.9 & 164.2& 163.8~\cite{HBando_2000}  \\ 
  Bi 4f$_{7/2}$ & 157.6  & 158.9&158.5~\cite{HBando_2000,TPDebies_1977} \\ \hline
 O 1s & - & 530.2&529.1~\cite{TPDebies_1977}, 529.9~\cite{HBando_2000} and 530.2~\cite{BVRChowdari_1996} \\ \hline
 C 1s & -& 285 & 284.5~\cite{HBando_2000} \\ \hline
\end{tabular}
\caption{XPS measured binding energies before and after exposure to ambient conditions. Extracted values are compared to previously reported XPS data. The observed values of the new peaks of Bi 4f and Te 3d are close to the binding energies for Bi$_2$O$_3$ and TeO$_2$ spectra, respectively.}
\label{table_1}
\end{table} 
\section{Shubnikov-de-Haas Oscillations}\label{SdH}
\begin{figure*}[!t]
\centering
\includegraphics[width=1.\textwidth]{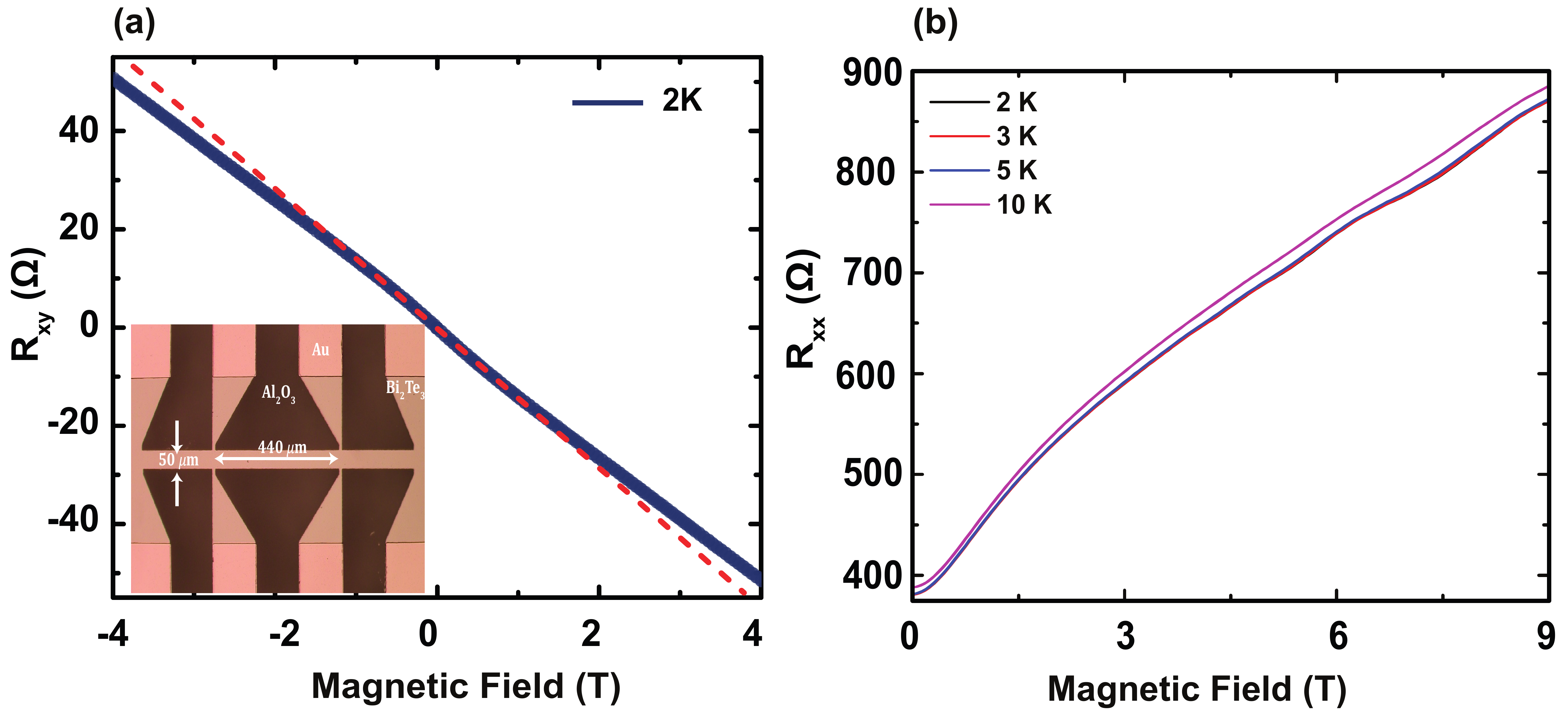}
\caption{\linespread{1.} (a) Nonlinear Hall effect in Bi$_2$Te$_3$ thin films. The insert shows typical optical micrograph image of a structured Hall bar device. Lateral dimensions of the Hall bar are indicated in the image. The ratio of length over width is: $L/W=8.8$. (b) Longitudinal resistance measured at various temperatures. Above a magnetic field of 5 T, small quantum oscillations are visible up to 10 K.}
 \label{fig:Second}
\end{figure*}

After growth and characterizations, samples were structured ex-situ into Hall bars by means of optical lithography and Ar ion beam etching (see insert Fig.~\ref{fig:Second}\textcolor{blue}{(a)}). Both the Hall and sheet resistances were measured as function of the applied magnetic field in the temperature range from 300 K down to 2 K. We used an excitation current of $1\mu$A in the measurements. 
\newline 
\newline
Figure~\ref{fig:Second}\textcolor{blue}{(a)} depicts the measured Hall resistance, $R_{xy}(B)$ at 2 K, plotted together with a one-carrier model fitting in order to highlight the presence of multiple bands carrier transport in $R_{xy}(B)$. The measured nonlinearity in $R_{xy}(B)$ suggests the presence of multiple carrier types. From the sign of the Hall signal, the majority charge carriers are found to be electrons. The  nonlinearity in $R_{xy}(B)$ has been measured in different TI materials and it was suggested to originate from the coexistence of bulk and surface transport channels \cite{Ren2010,Xiong2012,Bansal2012,Steinberg2010}. If all carriers participating in transport had the same mobility, $R_{xy}(B)$ would show linear behaviour with the slope determined by $1/(eR_H)$, where $R_H$ is the Hall coefficient. However, when there are multiple types of carriers with different (but comparable) mobilities, nonlinearity shows up in the $R_{xy}(B)$ data. From the low-field Hall coefficients ($R_H$) at 2K, using the procedure discussed in Ref.~\cite{PNgabonziza_2016}, we extracted a sheet carrier density of $n_{2D}=1/eR_H=5.8\times 10^{13}$ cm$^{-2}$ and mobility of $\mu=\frac{1}{e\,Rs\,n}= 2800$ cm$^2$ V$^{-1}$s$^{-1}$. If this sheet carrier density was interpreted as the carrier density of the topological surface state, with a Fermi velocity of about $3 \times 10^5$ m/s, then the chemical potential would be unreasonably high above the Dirac point. Bulk carriers are, therefore, the cause of the observed carrier density. From the Hall effect, we then extract a 3D carrier density of $n^{3D}_{_{Hall}}=n^{2D}_{_{Hall}}/d=8.2\times10^{18} $ cm$^{-3}$, with $d=70$ nm being the sample thickness. 
\newline 
\newline

\begin{figure*}[t]
\centering
\includegraphics[width=1.175\textwidth]{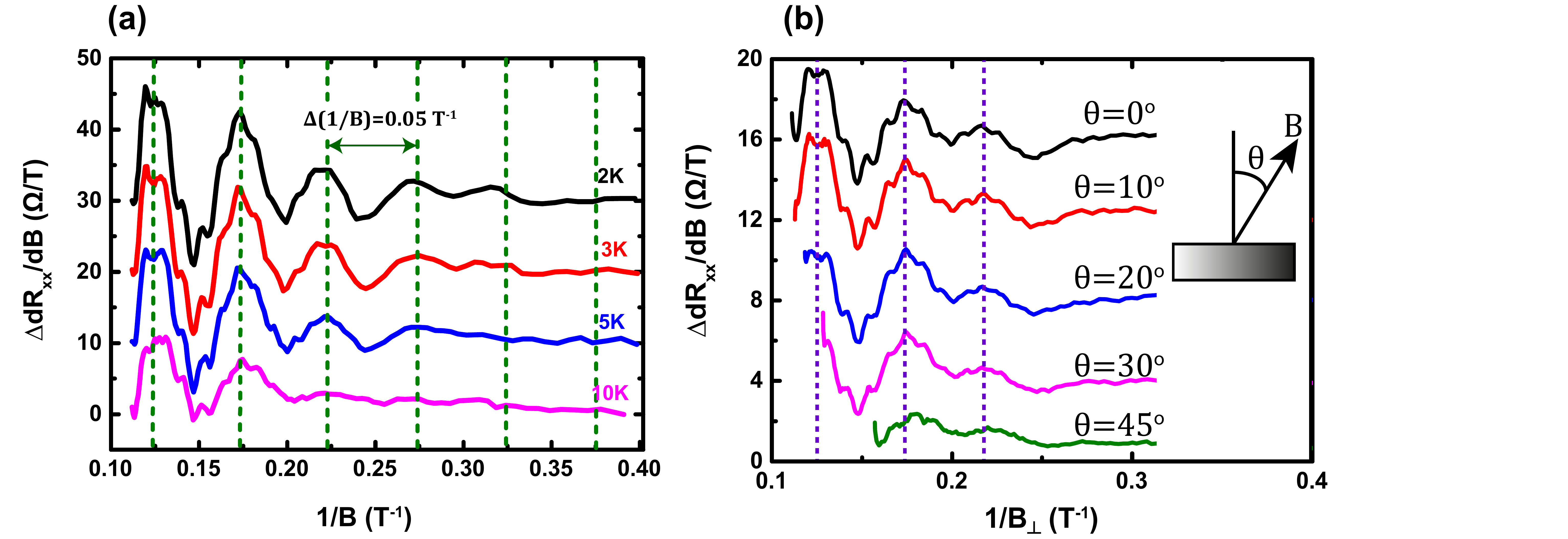}
\caption{\linespread{1.}(a) The oscillatory part of the longitudinal resistance $(\Delta dR_{xx}/dB)$ as function of $1/B$. After background subtraction, clear quantum oscillations are observed up to 10 K. (b) The SdH oscillations are plotted at different perpendicular component of applied magnetic field with $B_{\perp}=B\cos (\theta)$. Insert shows direction of applied magnetic field. For clarity, offsets are added in both graphs. }
 \label{fig:Third}
\end{figure*}

\begin{figure*}[!t]
\centering
\includegraphics[width=1.\textwidth]{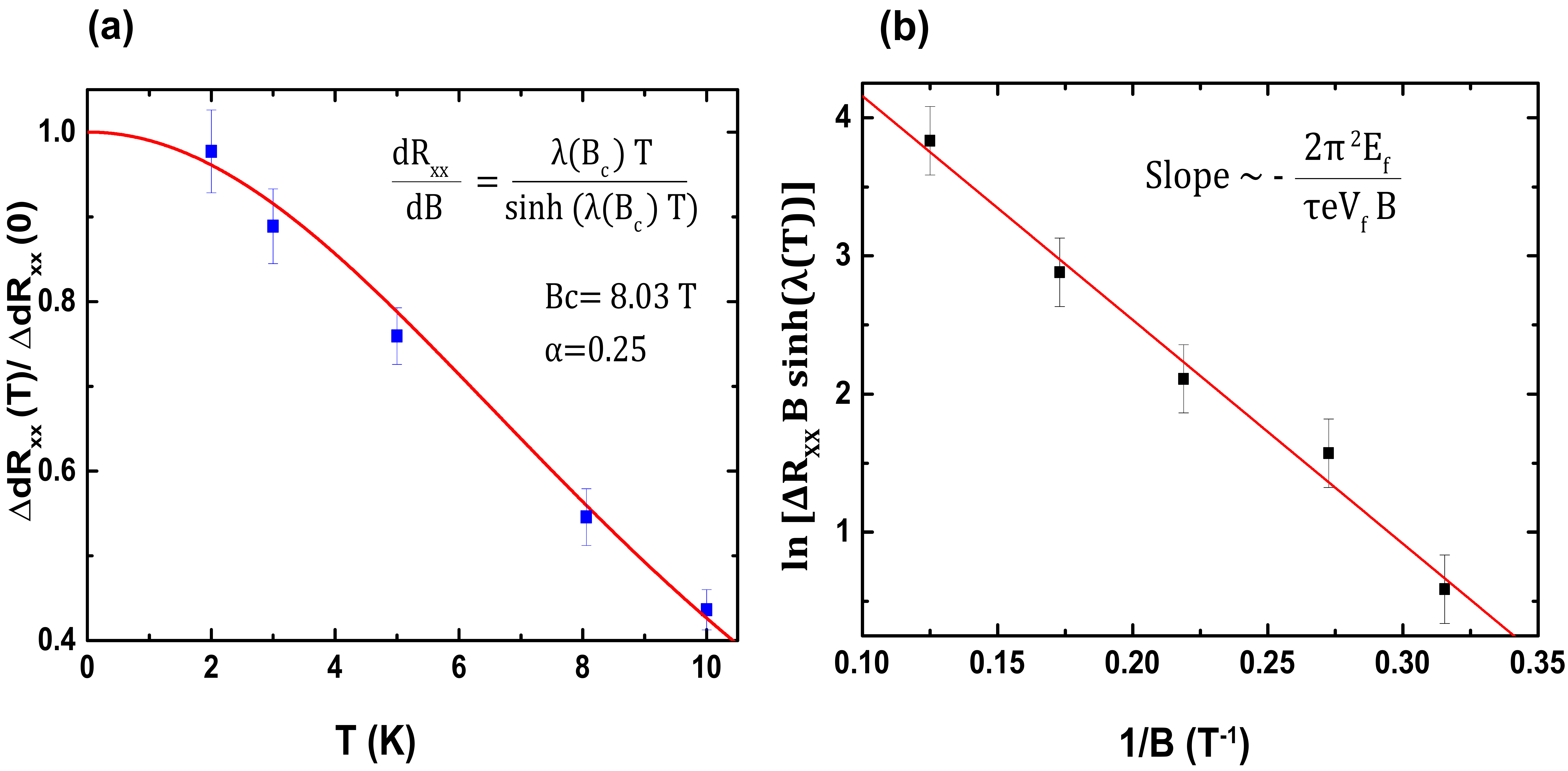}
\caption{\linespread{1.} (a)  Temperature dependence of the normalized longitudinal resistance. The solid red line is the best fit to the function $\lambda (T)/\sinh (\lambda(T))$. (b) The Dingle plot of $\ln [\Delta R_{xx}\,B \, \sinh (\lambda(T)]$ versus $1/B$ at 2 K. The transport lifetime $\tau$, mean free path $l$ and mobility $\mu$ are calculated from the parameters obtained from the best fit to this expression.}
 \label{fig:Fourth}
\end{figure*}

Next we focus on Shubnikov-de-Haas oscillations measured in Bi$_2$Te$_3$ thin films. Figure~\ref{fig:Second}\textcolor{blue}{(b)} depicts the longitudinal resistance, $R_{xx}$ as function of the applied perpendicular magnetic field for various temperatures. At high magnetic field ($B> 5$ T), $R_{xx}$ shows a superposition of small oscillations together with a quasi-linear increase up to a magnetic field of 9 T. The oscillations survive up to 10 K. The superimposed oscillations become clearly evident after the subtraction of a smooth polynomial background. Figure~\ref{fig:Third}\textcolor{blue}{(a)} gives the oscillatory part of the longitudinal resistance $(\Delta R_{xx})$, which displays periodic extrema with $1/B$. For the SdH oscillations at 2K, the extracted period of oscillations is $0.05\text{ T}^{-1}$. 
\newline 
\newline
If interpreted as oscillations from the carriers in a topological surface state with no spin degeneracy, the carrier density corresponding to this oscillation period would be $n=\frac{e}{•h}.\frac{1}{\Delta(1/B)}= 1.0\times 10^{12}$ cm$^{-2}$. This carrier density would give a Fermi wave vector of $k_F=2.5\times 10^6 \text{ cm}^{-1}$ and a Fermi energy only about 50 meV above the Dirac point, within the bulk band gap, which is clearly not consistent with the Hall data. 
\newline 
\newline
Figure~\ref{fig:Third}\textcolor{blue}{(b)} depicts the angle dependence of the SdH oscillations at 2 K. From these data we observe that periodic maxima and minima are overlapping when plotted against the perpendicular component of the magnetic field ($B_\perp = B \cos\theta$), as indicated by the dashed line in Fig.~\ref{fig:Third}\textcolor{blue}{(b)}. This scaling with angle suggests a two-dimensional nature of the oscillating channels. The 2D nature is often interpreted as coming from the TI surface states or from non-TI surface channels, like the trivial non-topological 2DEG states due to band-bending. Also the mixture of TI surface states and non-TI surface channels would also result into 2D nature of the oscillating channel~\cite{MBianchi_2010}. However, also the bulk Fermi surface can give rise to 2D quantum oscillations if it is cylindrical.
\newline 
\newline
Indeed, upon increasing the bulk charge carrier density, the FS becomes anisotropic such that $k_F$ along the $k_z$ direction gets considerably larger than $k_F$ in the $(k_x,k_y)$ plane. Thus, the FS will change from being a closed spherical FS at low carrier densities into an open cylinder-like FS at high carrier densities. A detailed discussion on the evolution of the FS is presented in Ref. \cite{ELahoud_2013}. Using the $k_F$ value extracted above from $n_{_{SdH}}^{2D}$ (2D carrier concentration from SdH oscillations), for a closed spherical FS the 3D carrier density would then be $n^{^{3D}}_{_{SdH}}=\frac{k_F^3}{3\pi}=1.6\times 10^{18}$cm$^{-3}$. On the contratry, if one considers a spin degenerate, cylindrical FS, where $k_z$ extends to the Brillouin zone boundary at $\pi/c$, the 3D carrier density is determined by $n^{^{3D}}_{_{SdH}}=\frac{\pi}{c}\, n^{^{2D}}_{_{SdH}}=1\times 10^{19}$cm$^{-3}$, with $c$ being the Bi$_2$Te$_3$ lattice parameter perpendicular to the surface. 
\newline 
\newline
Thus in comparing the Hall carrier density to the spherical and cylindrical FS carrier densities from SdH: $1.6\times 10^{18}$ cm$^{-3}<8.2\times10^{18}$cm$^{-3}<1\times 10^{19}$ cm$^{-3}$, these data show that the FS has a shape in between a spherical FS and open cylindrical FS, but more elongated towards a cylindrical one. This implies that the measured quantum oscillations come from bulk carriers, an observation that is in agreement with previous reports on thin films of the related material Bi$_2$Se$_3$ \cite{ELahoud_2013,EKVries_2017,HCao2012}. The two-dimensional nature of the oscillating channels in Fig~\ref{fig:Third} is consistent with the nearly cylindrical FS, rendering the conductance very anisotropic (but still bulk)~\cite{HCao2012}. 
\newline 
\newline
The mobility is determined through the analysis of SdH oscillations at different temperatures. The temperature-dependent amplitude of the oscillatory contribution to the resistance is described by the Lifshitz-Kosevich expression~\cite{LandauLifshitz1981}
\begin{equation}\label{LifshitzKosevich}
\Delta R \propto \sum_i e^{-\lambda_i T_{D_i}}\frac{\lambda_i (T)}{\sinh (\lambda_i (T) )} \sin \left( \frac{ 2\pi f_i}{B}+\phi_i  \right), 
\end{equation}
where $T$ is the temperature, $f_i$ and $\phi_i$ are the frequency and phase of the oscillations, respectively. $\lambda_i(T)$ is given by the expression: $\lambda_i (T)= 2\pi^2 k_BTm_{cycl}/(\hbar eB)$, with $m_{cycl}$, $\hbar$ and $k_B$ being the cyclotron mass, the reduced Planck constant and  the Boltzmann constant, respectively. The Dingle temperature, which has information about the quantum mobility, is given by: $T_{D_i}=\frac{2\pi^2 E_F}{\tau e B v_F^2}$, where $E_F$ is the Fermi energy, $\tau$ the transport life time, $e$ the electron charge and $v_F$ the Fermi velocity. At a constant magnetic field of $B_c$, the Lifshitz-Kosevich Eq. (\ref{LifshitzKosevich}) is simplified to
\begin{equation}\label{LifshitzKosevich_1}
\Delta R \propto \sum_i \frac{\alpha_i(B_c) T}{•\sinh (\alpha_i(B_c) T)},
\end{equation} 
with $\alpha_i= 2\pi^2 k_B m_{cycl}/(\hbar eB_c)$. Figure~\ref{fig:Fourth}\textcolor{blue}{(a)} gives the temperature dependence of the normalized longitudinal resistance oscillation amplitude extracted at a constant magnetic field $B_c=8.03$~T. Performing the best fit to Eq. (\ref{LifshitzKosevich_1}) yields a value of $\alpha=0.25$ from which $m_{cycl}$ is extracted to be $0.133\,m_e$ ($m_e$ is the free electron mass). In order to estimate the transport lifetime of the surface states, we use the Dingle plot~\cite{LHBao2012,MVeldhorst2012,Analytis2011} depicted in Fig.~\ref{fig:Fourth}\textcolor{blue}{(b)}. From the slope of the Dingle plot, we extract a transport lifetime of $\tau = 2.75\times 10^{-13}$~s, from which we derive the mean free path of $l=v_F \tau \simeq 150$~nm. Using the expression $\mu_s=e\tau/m_{cycl}=e\,l^{^{SdH}}/\hbar k_F$, the mobility is found to be $\mu_s\simeq 3600$ cm$^2$ V$^{-1}$s$^{-1}$, which is consistent with the mobility estimated from the Hall data.  An overview of the extracted charge-carrier properties for two Bi$_2$Te$_3$ thin films is given in Table~\ref{table_2}, revealing a consistent picture when the quantum oscillations are interpreted as due to high mobility bulk electrons. 
\section{Conclusion}
The consistency between bulk carrier densities and mobilities as determined from Hall and Shubnikov-de Haas data reveals that the bulk of the Bi$_2$Te$_3$ thin films dominates in transport. In single crystals, the charge carriers mainly arise from Te vacancies. Here, the thermodynamics of the MBE growth of thin film Bi$_2$Te$_3$, with a surplus of Te atoms, strongly reduces the number of Te vacancies, as shown by TEM. STEM microstructural analysis, on measured Hall bar devices, indicates an atomically sharp interfaces between the TI film and the substrate; and a continuous QL structure of the Bi$_2$Te$_3$ film. Only upon exposing the Bi$_2$Te$_3$ film surface to ambient conditions, the surface is chemically modified, as followed in time by XPS. Based on the fact that we know how the Fermi energy shifts into the bulk conduction band upon exposure to ambient conditions \cite{PNgabonziza_2015}, we speculate that the chemical surface modification effectively dopes electrons into the bulk without structural changes deeper inside the film. The observed bulk carrier density would be achieved already by an order of magnitude less than 1 electron per surface unit cell.
\newline 
\newline
In single crystals, a high bulk carrier density is the result of a large defect density. The defects and vacancies act as scattering centers too. An increased carrier density thereby results in a lower bulk mobility. The low amount of Te vacancies in our thin films provides record high mobility values (for bulk carriers in topological thin films). With these high values for the bulk mobility, Shubnikov-de Haas oscillations appear in the magnetotransport of the bulk conduction channel. Because of the cylindrical shape of the Fermi surface, the angle dependence of the bulk magnetoresistance oscillations are two-dimensional in nature (not being due to the surface states but due to anisotropic bulk transport). With the advance in topological insulator thin film quality we foresee that these findings will be relevant for many applications, as electric field gating can also populate bulk bands in films that are capped and intrinsically insulating in the bulk \cite{PNgabonziza_2016}. Very recently single crystal growers have modified the Bi$_2$Te$_3$ growth by applying an additional `defect cleaning' step, by which Te vacancies are also strongly reduced, providing very large bulk mobilities also in this case~\cite{Shrestha_2017}.

\begin{table}[t]
\centering
\begin{tabular}{|p{6.5 cm}|p{4.3 cm}|p{4.3 cm}|p{}}
\hline
 \textbf{Extracted from } & \textbf{Mobility} [cm$^2$.v$^{-1}$.s$^{-1}$]&\textbf{Carrier density} [cm$^{-3}$] \\ \hline
  Hall signal at a temperature of 2K &  $\mu^{S_1}=2800$ & $n^{^{3D}}_{S_1}=8.2\times 10^{18}$ \\
      & $\mu^{S_2}=2300$  & $n^{^{3D}}_{S_2}=9.8\times 10^{18}$\\  \hline
 Shubnikov-de-Haas Oscillations&  $\mu^{S_1}=3600$ & $n^{^{3D}}_{S_1}=1.0\times 10^{19}$\\ 
 & $\mu^{S_2}=3000$  & $n^{^{3D}}_{S_2}=1.3\times 10^{19}$\\  \hline
  
\end{tabular}
\caption{Overview of the extracted transport characteristics (mobility and carrier density, from Hall signal and Shubnikov-de-Haas oscillations). Data are for two Bi$_2$Te$_3$ thin films ($S_1$ of thickness 70 nm and $S_2$ of thickness 50 nm). For SdH oscillations, only cylindrical FS 3D carrier densities are presented.}
\label{table_2}
\end{table} 
\section*{Acknowledgement}
This work was financially supported by the Netherlands Organization for Scientific Research (NWO) and the European Research Council (ERC) through a Consolidator Grant.

%-----------------------------------------------------------------------------

\newpage
\begin{thebibliography}{99}
\bibitem{CLKane_2005}C. L. Kane and E. J. Mele, Phys. Rev. Lett. \textbf{95}, 226801 (2005).
\bibitem{LFu_2007} L. Fu, C. L. Kane, and E. J. Mele, Phys. Rev. Lett.\textbf{ 98}, 106803 (2007).
\bibitem{CWJBeenakker_2013} C. W. J. Beenakker, Annu. Rev. Condens. Matter Phys. \textbf{4}, 113 (2013).
\bibitem{YYLi_2010}  Y. Y. Li, G. A. Wang, X. G. Zhu, M. H. Liu, C. Ye, X. Chen, Y. Y. Wang, K. He, L. L. Wang, X. C. Ma, H. J. Zhang, X. Dai, Z. Fang, X. C. Xie, Y. Liu, X. L. Qi, J. F. Jia, S. C. Zhang, and Q. K. Xue, Adv. Mater. \textbf{22}, 4002 (2010).
\bibitem{MBrahlek_2015}M. Brahlek, N. Koirala, N. Bansal, S. Oh, Solid State Communications \textbf{215}, 54 (2015).
\bibitem{NBansal_2012} N. Bansal, Y. S. Kim, M. Brahlek, E. Edrey, and S. Oh, Phys. Rev. Lett. \textbf{109}, 116804 (2012).
\bibitem{PDCking_2014} P. D. C. King, R. C. Hatch, M. Bianchi, R. Ovsyannikov, C. Lupulescu, G. Landolt, B. Slomski, J. H. Dil, D. Guan, J. L. Mi, E. D. L. Rienks, J. Fink, A. Lindblad, S. Svensson, S. Bao, G. Balakrishnan, B. B. Iversen, J. Osterwalder, W. Eberhardt, F. Baumberger, and Ph. Hofmann, Phys. Rev. Lett. \textbf{107}, 096802 (2014).
\bibitem{MBianchi_2010} M. Bianchi, D. Guan, S. Bao, J. Mi, B. B. Iversen,	P. D.C. King, and P. Hofmann, Nat. Comm. \textbf{1}, 128 (2010).
\bibitem{Ren2010}Z. Ren, A. A. Taskin, S. Sasaki, K. Segawa and Y. Ando, Phys. Rev. B, \textbf{82} 241306(R) (2010).
\bibitem{DXQu2010_Science} D. -X. Qu, Y. S. Hor, J. Xiong, R. J. Cava, N. P. Ong, Science, \textbf{329} 5993 (2010).
\bibitem{KShrestha_2014PRB} K. Shrestha, V. Marinova, B. Lorenz, and P. C. W. Chu, Phys. Rev. B, \textbf{90}  241111(R) (2014).
\bibitem{PNgabonziza_2015}P. Ngabonziza, R. Heimbuch, N. de Jong, R. A. Klaassen, M. P. Stehno, M. Snelder, A. Solmaz, S. V. Ramankutty, E. Frantzeskakis, E. van Heumen, G. Koster, M. S. Golden, H. J. W. Zandvliet, and A. Brinkman, Phys. Rev. B \textbf{92}, 035405 (2015).
\bibitem{PNgabonziza_2016} P. Ngabonziza, M. P. Stehno, H. Myoren, V. A. Neumann, G. Koster, and A. Brinkman, Adv. Electron. Mater. \textbf{2}, 1600157  (2016).
\bibitem{TChen_2014}T. Chen, Q. Chen, K. Schouteden, W. Huang, X. Wang, Z. Li, F. Miao, X. Wang, Z. Li, B. Zhao, S. Li, F. Song, J. Wang, B. Wang, C. Van Haesendonck, and G. Wang, Nat. Commun. \textbf{75}, 5022 (2014).
\bibitem{MBrahlek_2014}M. Brahlek, N. Koirala, M. Salehi, N. Bansal, and S. Oh, Phys. Rev. Lett. \textbf{113}, 026801 (2014).
\bibitem{MLanius_2016} M. Lanius, J. Kampmeier, C. Weyrich, S. Ko\"{o}lling, M. Schall, P. Sch\"{u}ffelgen, E. Neumann, M. Luysberg, G. Mussler, P. M. Koenraad, T. Sch\"{a}pers, and D. Gr\"{u}tzmacher, Cryst. Growth Des. \textbf{16}, 2057 (2016).
\bibitem{MEschbach_2016} M. Eschbach, E. M\l{}y\'{n}´czak, J. Kellner, J. Kampmeier, M. Lanius, E. Neumann, C. Weyrich, M. Gehlmann, P. Gospodaric, S. D\"{o}ring, G. Mussler, N. Demarina, M. Luysberg, G. Bihlmayer, T. Sch\"{a}pers, L. Plucinski, S. Bl\"{u}gel, M. Morgenstern, C. M. Schneider, and D. Gr\"{u}tzmacher, Nat. Comm. \textbf{6}, 8816 (2015).
\bibitem{JWang_2012}J. Wang, X. Chen, B. -F. Zhu, and S. -C. Zhang, Phys. Rev. B \textbf{85}, 235131 (2012).
\bibitem{JZhang_2011} J. Zhang, C. Z. Chang, Z. Zhang, J. Wen, X. Feng, K. Li, M. Liu, K. He, L. Wang, X. Chen, Q. K. Xue, X. Ma, Y. Wang,     Nat.  Commun. \textbf{2}, 574 (2011).
\bibitem{YXu_2014} Y. Xu, I. Miotkowski, C. Liu,	J. Tian,	H. Nam,	N. Alidoust, J. Hu, C. -K. Shih, M. Z. Hasan, and Y. P. Chen, Nat. Phys. \textbf{10}, 956 (2014).
\bibitem{TArakane_2012} T. Arakane, T. Sato, S. Souma, K. Kosaka, K. Nakayama, M. Komatsu, T. Takahashi, Z. Ren,K. Segawa, and Y. Ando, Nat. Commun. \textbf{3}, 636 (2012).
\bibitem{KHoefer_2015} K. Hoefer, C. Becker, S. Wirth, and L. H. Tjeng, AIP Adv. \textbf{5}, 097139 (2015).
\bibitem{ZAlpichshev_2010}
Z. Alpichshev, J. G. Analytis, J. -H. Chu, I. R. Fisher, Y. L. Chen, Z. X. Shen, A. Fang, A. Kapitulnik, Phys. Rev. Lett. \textbf{104}, 016401 (2010).
\bibitem{MVeldhorst_2013}M. Veldhorst , M. Snelder, M. Hoek, C. G. Molenaar, D. P. Leusink, A. A. Golubov, H. Hilgenkamp and A. Brinkman, Phys. Status Solidi RRL  \textbf{7}, 1 (2013), and the references therein.
\bibitem{ELahoud_2013} E. Lahoud, E. Maniv, M. Shaviv Petrushevsky, M. Naamneh, A. Ribak, S. Wiedmann, L. Petaccia, Z. Salman, K. B. Chashka, Y. Dagan and A. Kanigel, Phys. Rev. B \textbf{88}, 195107 (2013).
\bibitem{AATaskin_2012} A. A. Taskin, S. Sasaki, K. Segawa and Y. Ando. Phys. Rev. Lett. \textbf{109}, 066803 (2012).
\bibitem{LHBao2012} L. H. Bao, L. He, N. Meyer, X. F. Kou, P. Zhang, Z. G. Chen, A.V. Fedorov, J. Zou, T. M. Riedemann, T. A. Lograsso,K. L.Wang, G. Tuttle and F. X. Xiu, Sci. Rep. \textbf{2}, 726 (2012).
\bibitem{MVeldhorst2012}M. Veldhorst, M. Snelder, M. Hoek, T. Gang, V. K. Guduru, X. L. Wang, U. Zeitler, W. G. van der Wiel, A. A. Golubov, H. Hilgenkamp and A. Brinkman, Nature Mater. \textbf{11}, 417 (2012).
\bibitem{KWang_2013}K. Wang, Y. Liu, Weiyi Wang, N. Meyer, L. H. Bao, L. He, M. R. Lang, Z. G. Chen, X. Y. Che, K. Post, J. Zou, D. N. Basov, K. L.Wang, and F. Xiu, Appl. Phys. Lett. \textbf{103}, 031605 (2013).
\bibitem{Petrushevsky_2012} M. Petrushevsky, E. Lahoud, A. Ron, E. Maniv, I. Diamant, I. Neder, S. Wiedmann, V. K. Guduru, F. Chiappini, U. Zeitler, J. C. Maan, K. Chashka, A. Kanigel, and Y. Dagan, Phys. Rev. B \textbf{86}, 045131 (2012).
\bibitem{Devidas_2014}T. R. Devidas, E. P. Amaladass, S. Sharma, R. Rajaraman, D. Sornadurai, N. Subramanian, A. Mani, C. S. Sundar, and A. Bharathi, Europhys. Lett. \textbf{108}, 67008 (2014).
\bibitem{CZhang_2014}C. Zhang, X. Yuan, K.Wang, Z.-G. Chen, B. Cao,W.Wang, Y. Liu, J. Zou, and F. Xiu, Adv. Mater. \textbf{26}, 7110 (2014).
\bibitem{PNgabonziza_2018} P. Ngabonziza, M. P. Stehno, G. Koster, and A. Brinkman; \textit{In-situ Characterization Tools for Bi$_2$Te$_3$ Topological Insulator Nanomaterials}, Book chapter in the Springer book series on ``\textit{In-situ Characterization Techniques for Nanomaterials}"; Challa S. S. R. Kumar (ed.) Springer-Verlag Berlin Heidelberg, ISBN: 978-3-662-56322-9 (2018). 
\bibitem{LHe_2013}L. He, X. Kou and K. L. Wang, Phys. Status Solidi RRL \textbf{7}, 50 (2013) and reference therein.
\bibitem{XLiu_2011}X. Liu, D. J. Smith, J. Fan, Y. -H. Zhang, H. Cao, Y. P. Chen, J. Leiner, B. J. Kirby, M. Dobrowolska, and J. K. Furdyna, Appl. Phys. Lett. \textbf{99}, 171903 (2011).
\bibitem{AKoma_1992} A. Koma, Thin Solid Films \textbf{216}, 72 (1992).
\bibitem{HBando_2000}
H. Bando, K. Koizumi, Y. Oikawa, K. Daikohara, V. A. Kulbachinskii and H. Ozaki, J. Phys.: Condens. Matter \textbf{12}, 5607 (2000). 
\bibitem{BVRChowdari_1996} B. V. R Chowdari  and P. P Kumari, J. Non-Cryst. Solids \textbf{197}, 31 (1996).
\bibitem{TPDebies_1977} T. P Debies and J. W Rabalais, Chem. Phys. \textbf{20}, 277 (1977).
\bibitem{AJGreen_2016} A. J. Green, S. Dey, Y. Q. An, B. O'Brien, S. O'Mullane, B. Thiel, and A. C. Diebold, J. Vac. Sci. Technol. A \textbf{34}, 6 (2016).
\bibitem{Xiong2012} J. Xiong, Y. Luo, Y. Khoo, S. Jia, R. J. Cava and N. P. Ong, Phys. Rev. B \textbf{86}, 045314 (2012).
\bibitem{Bansal2012} N. Bansal, Y. S. Kim, M. Brahlek, E. Edrey and S. Oh, Phys. Rev. Lett. \textbf{109}, 116804 (2012).
\bibitem{Steinberg2010}H. Steinberg, D. R. Gardner, Y-S. Lee, P. Jarillo-Herrero, Nano Lett. \textbf{10}, 5032 (2010).
\bibitem{EKVries_2017} E. K. de Vries, S. Pezzini, M. J. Meijer, N. Koirala, M. Salehi, J. Moon, S. Oh, S. Wiedmann, and T. Banerjee, Phys. Rev. B \textbf{96}, 045433 (2017). 
\bibitem{HCao2012}H. Cao, J. Tian, I. Miotkowski, T. Shen, J. Hu, S. Qiao and Y. P. Chen, Phys. Rev. Lett. \textbf{108}, 216803 (2012).
\bibitem{LandauLifshitz1981}L. D. Landau and E. M. Lifshitz. \textit{Course of theoretical physics.}, 1\textsuperscript{st} Edition, Robert Maxwell, M. C., Oxford-UK,  (1981).
\bibitem{Analytis2011}J. G. Analytis, R. D. McDonald,	S. C. Riggs, J.-H. Chu, G. S. Boebinger and I. R. Fisher, Nat. Phys. \textbf{6}, 960 (2010).
\bibitem{Shrestha_2017} K. Shrestha, M. Chou, D. Graf, H. D. Yang, B. Lorenz, C. W. Chu, Phys. Rev. B \textbf{95}, 195113 (2017).
\end{thebibliography}
\end{document}